\documentclass{pasa}%

\def\gsim{ \lower .75ex \hbox{$\sim$} \llap{\raise .27ex \hbox{$>$}} }
\def\lsim{ \lower .75ex\hbox{$\sim$} \llap{\raise .27ex \hbox{$<$}} }

\def\beq{\begin{equation}}
\def\eeq{\end{equation}}

\def\sw{{\it Swift}}
\def\fe{{\it Fermi}}
\def\ba{BATSE}
\def\cgro{{\it CGRO}}

\def\eiso{$E_{\rm iso}$}
\def\ama{$E_{\rm p}-E_{\rm iso}$}

\def\th{$\theta_{\rm jet}$}

\def\thv{$\theta_{\rm view}$}
\def\tjet{$t_{\rm break}$}
\def\tpeak{$t_{\rm peak}$}
\def\egamma{$E_{\gamma}$}

\def\G{$\Gamma_{0}$}

\def\egcom{$E'_{\gamma}$}

\def\epcom{$E'_{\rm p}$}

\title[GRB orphan afterglows -- Radio transients]{GRB orphan afterglows in present and future radio transient surveys}
\author[Ghirlanda et al.]{G. Ghirlanda$^{1}$\thanks{giancarlo.ghirlanda@brera.inaf.it}, D. Burlon$^{2,3}$, G. Ghisellini$^1$, R. Salvaterra$^4$, M. G. Bernardini$^1$,
S. Campana$^1$,\\ S. Covino$^1$, P. D'Avanzo$^1$,V. D'Elia$^{5,6}$, A. Melandri$^1$, T. Murphy$^{2,3}$, L. Nava$^7,8$, S. D. Vergani$^{9}$, \\ 
\and  G. Tagliaferri$^1$\\
\\
\affil{$^1$INAF -- Osservatorio Astronomico di Brera, via E. Bianchi 46, I-23807 Merate (LC) - Italy}%
\affil{$^2$Sydney Institute for Astronomy, The University of Sydney, NSW 2006, Australia}%
\affil{$^3$ARC Centre of Excellence for All-sky Astrophysics (CAASTRO)}%
\affil{$^4$INAF -- IASF Milano, via E. Bassini 15, I-20133 Milano, Italy }%
\affil{$^{4}$ASI -- Science Data Center, via Galileo Galilei, 00044 Frascati, Italy}
\affil{$^{5}$INAF-OAR Via Frascati 33, I-00040 Monteporzio Catone Italy}
\affil{$^7$APC Universit\'e Paris Diderot, 10 rue Alice Domon et Leonie Duquet, F-75205 Paris Cedex 13, France }%
\affil{$^8$Racah Institute of Physics, The Hebrew University of Jerusalem, 91904, Israel }%
\affil{$^9$GEPI, Observatoire de Paris, CNRS, Univ. Paris Diderot, 5 place Jules Janssen, 92190, Meudon, France}}%
\jid{PASA}
\doi{10.1017/pas.\the\year.xxx}
\jyear{\the\year}

\begin{document}%
\begin{abstract}
Orphan Afterglows (OA) are slow transients produced by Gamma Ray Bursts seen off--axis that become visible on timescales of 
days/years at optical/NIR and radio frequencies, when the prompt emission at high energies (X and $\gamma$ rays) has already ceased. 
Given the typically estimated jet opening angle of GRBs \th$\sim 3^{\circ}$,  
for each burst pointing to the Earth there should be a factor $\sim700$  more GRBs pointing in other directions. 
Despite this, no secure OAs have been detected so far. Through a population synthesis code we study the 
emission properties of the population of OA at radio frequencies. OAs reach their emission peak on year-timescales and they last for a 
comparable amount of time. The typical peak fluxes (which depend on the observing frequency) are of few $\mu$Jy in the radio band 
with only a few OA reaching the mJy level. These values are consistent with the upper limits on the radio flux  of SN Ib/c observed at late times. We find that the OA radio number count distribution has  a typical slope $-1.7$ 
at high fluxes and a flatter ($-0.4$) slope at low fluxes with a break at a frequency--dependent flux. Our predictions of the OA rates are 
consistent with the (upper) limits of recent radio surveys and archive searches for radio transients. Future radio surveys like 
VAST/ASKAP at 1.4 GHz should detect $\sim3\times10^{-3}$ OA deg$^{-2}$ yr$^{-1}$, MeerKAT and EVLA  at 8.4 
GHz should see $\sim3\times10^{-1}$ OA deg$^{-2}$ yr$^{-1}$. The SKA, reaching the $\mu$Jy flux limit, could see up to 
$\sim0.2-1.5$ OA deg$^{-2}$ yr$^{-1}$. These rates also depend on the duration of the OA above a certain flux 
limit and we discuss this effect with respect to the survey cadence. 
\end{abstract}
\maketitle%

\section{Introduction}
In the standard external shock model of Gamma Ray Bursts (GRBs) the afterglow emission is produced when the ultra relativistic jet 
is decelerated by the interstellar medium (Meszaros \& Rees 1997). During this phase the bulk Lorentz factor $\Gamma$  decreases 
with time while the beaming angle of the emitted radiation $\Omega(t)\propto 1/\Gamma(t)^{2}$ increases.  
Moreover, there is a large consensus both for theoretical and observational reasons, that GRBs are jetted sources.
The estimate of the jet opening angle can be derived from the time of the afterglow light curve steepening (the jet break time \tjet) due to the Lorentz factor becoming $\sim$1/\th\ (Rhoads et al. 1997).  Typical jet opening angles derived from the afterglow light curve breaks  (Frail et al. 2001; Ghirlanda et al. 2004) are clustered around \th=0.05 radiants. 

Therefore,
a GRB which is observed off--axis, with a viewing angle \thv$>$\th, will be undetected as a prompt burst of $\gamma$--ray 
photons because the prompt emission,  produced by material moving with a bulk Lorentz factor $\Gamma_0\sim 10^{2}-10^{3}$, is beamed 
within an angle 1/\G$<$\th. However, the  afterglow emission (in the optical/NIR and radio band) can be detected when the beaming angle 
of the radiation intercepts the line of sight, i.e.  1/$\Gamma \sim$\thv. After this time the emission for an off-axis observer is the same 
that would be seen if \thv$<$\th. 


Orphan Afterglows (OA) are GRBs seen off-axis detectable at any frequency, without the high energy 
$\gamma$--ray counterpart. For this reason their study follows a different path than  normal GRBs (where it is the high energy trigger to initiate a follow up campaign to monitor the afterglow emission at different frequencies). OAs can be detected as transients through wide field deep surveys and they could be a considerable fraction of the population of detected transients (Rohads 1997, 2003;  Nakar, Piran \& Granot 2002; Totani \& Panaitescu 2002).  Considering the typical jet opening angles measured in GRBs (e.g. Frail et al. 2001; Ghirlanda et al. 2007) \th$\sim 3^{\circ}$, for each GRB detected in the $\gamma$--ray band  there should be $\sim730$ (i.e. $\sim2/$\th$^2$) bursts pointing in any other direction. These are orphan 
afterglows.

Upper limits \th$\lsim$22$^\circ$ (Levinson et al. 2002) or lower limits  \th$\gsim$0.8$^\circ$ (Soderberg et al. 2006) on the typical opening angle of GRBs were inferred from OA searches. 
However, to date no orphan afterglow has been confirmed. Some transients have been identified as possible OA candidates through archival searches both at 
radio frequencies (e.g. see Murphy et al. 2013; Bell et al. 2011 for a summary of radio OA search results) and in the optical band (e.g. Rau, 
Greiner \& Schwarz  2006; Malacrino et al. 2007; Zou, Wu \& Dai  2007) but none of these has been confirmed as an orphan GRB afterglow.  
Even the very recent discovery (Cenko et al. 2013) of an optical transient by the Palomar Transient Survey (PTF) seems favor a ``dirty fireball" or an untriggered bursts origin (i.e. both scenarios related to a GRB pointing towards the Earth) rather than to an orphan afterglow (Cenko et al. 2013).

A possible strategy (Soderberg et al. 2006; Bietenholz et al. 2013) for identifying off--axis GRBs is to observe a considerable radio emission from type Ib/c SNe at late times (i.e. years after the SN explosion). The estimate of the size of the SN event at late epochs allows us to verify if the SN had a relativistic jet. A handful of SNe  were found to have a bright radio emission: SN 2001em - (Granot \& Ramirez Ruiz 2004), SN 2007gr (Paragi et al. 2010) and SN 2009bb (Soderberg et al. 2010), SN 2003gk (Bietenholz et al. 2013). However either subsequent revision of the radio observations (Soderberg et al. 2010a) or VLBI observations (Bietenholz, BArtel \& Rupen 2010; Bietenholz et al. 2013) showed that these events are non--relativistic SN Ib/c. Only SN 2009bb could be considered as a transition event between the class of SNe and GRBs (Soderberg et al. 2010a).

If GRBs have a jet, OA should exist and they should be detected by 
wide field deep surveys. Here we consider the standard model of a uniform jet for GRBs but the detection rate of OA has been explored (Rossi, 
Perna \& Daigne 2008) also for a universal structured jet model (Rossi, Lazzati \& Rees 2002; Zhang \& Meszaros 2002). 
\begin{figure}
\includegraphics[width=8.5cm,trim=20 10 20 10,clip=true]{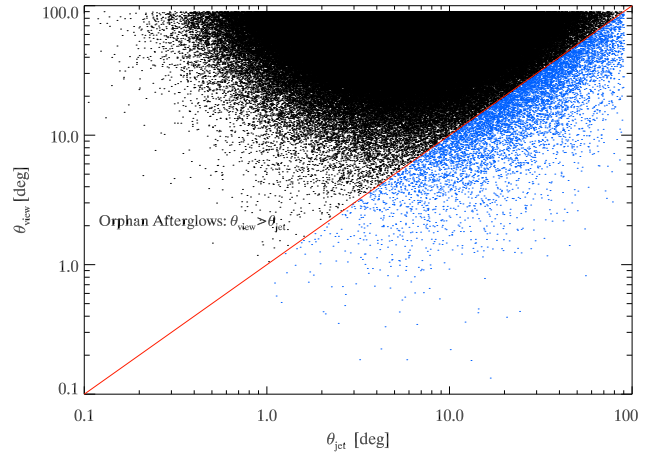}
\caption{Viewing angle (\thv) versus jet opening angle (\th) of the simulated population of GRBs (G13). The solid line of equality separates GRBs 
pointing to the Earth (blue symbols - with \th$\ge$\thv) from the bulk  of the population of GRBs not pointing to the Earth (black symbols - with \th$\le$\thv) 
which can be detected as Orphan Afterglows. }
\label{fg1}
\end{figure}

If the  non detection of OA is due to the low sensitivity of past surveys, future deep and wide field transient surveys could succeed in detecting OA and,  as mentioned above, they could represent a considerable fraction of the detected transient population  
(e.g. Frail et al. 2012).
However, one  challenge is how to 
disentangle the population of GRB OA from other possible sources producing similar transients. In the radio band, the detection of GRB afterglows 
(Chandra \& Frail 2012) benefits from the fact that the brightest phase of the emission happens on timescales of few days (for GRBs observed on--axis), 
because initially suppressed by self--absorption. This allows us to detect and follow the GRB radio emission when the flux at higher frequencies is 
already below the  sensitivity of  available instruments. Late time observations of radio emission from GRBs is fundamental for constraining some key 
parameters of these sources (e.g. Berger, Kulkarni \&  Frail 2003, 2004) and for estimating their global energetic (Frail, Kulkarni \& Nicastro 1997; 
Frail, Metzger \& Berger 2004; Shivvers \& Berger 2011; Sironi \& Giannios 2013). 

We are now entering a new era of radio surveys. In preparation for the Square Kilometre Array (SKA, Carilli \& Rawlings, 2004), that will represent 
a giant leap forward in survey depth at all GHz and sub-GHz frequencies, there are several pathfinders that will become operative within the next 
few years. In the sub-GHz regime, both the Low Frequency Array for Radio-astronomy (LOFAR, van Haarlem et al. 2013) and the Murchinson Widefield 
Array (MWA, Bowman et al. 2013, Tingay et al. 2013) will scan the sky with unprecedented survey speed, thanks to their field of view. Nonetheless 
as far as the search for OA is concerned, these will be likely limited to the exceptionally bright sources. In the GHz band, the most promising telescopes -in 
terms of dynamic range and fields of view- for the systematic search for transients are the Australian SKA-Pathfinder (ASKAP, Johnston et al. 2007), the 
Aperture Tiles In Focus (AperTIF) experiment, a wide field upgrade to the Westerbrok Synthesis Radio Telescope (Verheijen et al. 2008), and the MeerKAT 
telescope. While the MHz telescopes are already operative, the GHz SKA pathfinders will take first light in the next 2--5 years. 

In order to study the detectability of OA with on--going and future surveys it is necessary to know the emission properties (timescales and flux level) of 
the population of OA which depend on the properties (energetics, distance scale, jet opening angle) of the population of GRBs. In this paper we use a 
population synthesis code PSYCHE (Ghirlanda et al. 2013a; 2013b)  summarised in \S2 which has already been used to predict the detectability of 
on--axis GRB afterglows in the radio band (Ghirlanda et al. 2013b). Here we explore for the first time with such a code the radio emission properties 
of Orphan Afterglows (\S3) and compare with the current limits on the detection rates at radio frequencies (\S4). We explore the detectability of OA by 
future radio surveys in \S5.

\section{The population of Gamma Ray Bursts}

Ghirlanda et al. (2013a - G13 hereafter) built a population synthesis code which simulates GRBs (i) distributed in the Universe up to $z=10$ according 
to the GRB formation rate (Salvaterra et al. 2012), (ii) with initial bulk Lorentz factors \G\  and (iii) jet opening angles \th\ extracted from log--normal distributions. 
Each burst is oriented with respect to the line of sight with a viewing angle \thv\ (distributed as the $\sin$ \thv\ probability density function). The starting 
assumption of PSYCHE is that all GRBs have a standard comoving frame energy  \egcom$=1.5\times10^{48}$ erg and a unique comoving frame prompt 
emission peak energy \epcom$=1.5$ keV. This assumption is motivated by the  clustering of these quantities found when correcting for the beaming factor \G,  
estimated from the peak of the afterglow light curve (Liang et al. 2010; Ghirlanda et al. 2012; Lu et al. 2012). 

The energetic of each simulated burst is then 
determined by \G\ (\egamma=\G\egcom) and the isotropic equivalent energy by \G\ and \th\ (\eiso=\G\egcom/(1-cos\th)). The simulation free parameters (e.g. 
the parameters of the log--normal distributions of \G\ and \th) are determined by reproducing the observed properties of the GRB samples observed by 
different satellites: (a) the flux distribution and (b) the empirical \ama\ correlation of BAT6, i.e. a complete sample of bright \sw\ GRBs  (Salvaterra et al. 2012; 
Nava et al. 2012), (c) the flux and fluence distributions of GRBs detected by \fe\ and by \ba-\cgro. In particular, constraint (a) is used to  normalise the GRB 
population so that the number of simulated bursts with peak flux $>$2.6 ph cm$^{-2}$ s$^{-1}$ matches the rate of GRBs ($\approx$15 sr$^{-1}$ yr$^{-1}$) 
detected by \sw-BAT above this flux threshold. 

The GRBs used as observational constraints by PSYCHE are bursts that are pointing towards the Earth (i.e. with \th$>$\thv), indeed their prompt emission 
has been detected by different satellites (\sw, \fe, \cgro). However, the  code simulates also GRBs seen off--axis. This is because one of the scopes of G13 
was to describe the parent population of bursts, of which only a minor fraction (with \th$>$\thv) can be detected in the $\gamma$--ray band. The simulated 
population of GRBs is shown in Fig.\ref{fg1} where the viewing angle \thv\ is plotted against the jet opening angle \th. It is evident that the jet angle distribution 
\th\ is log--normal, while the clustering of the population towards large \thv\ values is the effect of the probability function of \thv. 

The population is composed by a minor fraction ($\sim$2.4\%) of GRBs that are ``pointing" towards the Earth (i.e. with \th$>$\thv) which can be detected 
by $\gamma$--ray detectors (blue symbols in Fig.\ref{fg1}) and a majority of bursts that are off--axis with \th$<$\thv\ (black symbols in Fig.\ref{fg1}). The latter 
are events that can be potentially detected at any frequency except that in the $\gamma$--ray band and are called OA. Therefore, a description of the properties 
of the entire GRB population should also consider the GRBs that are off--axis and are detectable only as OA. According to the results of PSYCHE, the ratio of 
the off--axis to on--axis bursts (i.e. black vs. blue dots in Fig.\ref{fg1}) is $\sim$40, smaller than the typical value obtained by assuming that all GRBs have 
\th=3$^{\circ}$. This is due to the distribution of \th\ that we find with PSYCHE to be a log--normal (see G13 for details). Through our simulation we find that 
the rate of  OA is $\sim3.3\times10^4$ yr$^{-1}$ sr$^{-1}$ (i.e. $\sim$10 yr$^{-1}$ deg$^{-2}$). 


\section{Orphan afterglow timescales}

One of our aims is to derive the detectability of OAs with current and future radio surveys. We consider here the flux at the brightest phase of the OA 
emission. This happens at a characteristic time $t_{\rm v}$ when the bulk Lorentz factor $\Gamma(t_{\rm v})=1/\sin\theta_{\rm view}$. Although the flux 
starts to rise before this time, when the edge of the jet (that we assume here to have a sharp top-hat conical section) closer to the viewing angle becomes 
visible (i.e. when $\Gamma=1/\sin$(\thv-\th)), the peak flux of the orphan afterglow happens when all the jet is visible. For $t\ge t_{\rm v}$ the light curve 
as seen by an observer off--axis is the same as that seen by an on--axis observer. 
\begin{figure}
\includegraphics[width=8.5cm,trim=40 10 20 10,clip=true]{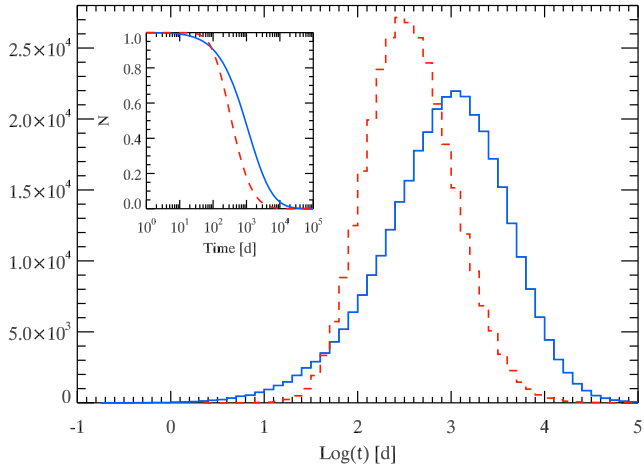}
\caption{Differential and cumulative (inset) distributions of the time when orphan afterglows peak (blue solid line) and of the duration of 
the orphan afterglow emission (red dashed line).
}
\label{fg2}
\end{figure}

The peak of the OA emission is reached when the afterglow is still described by the relativistic Blandford-McKee (1976) self similar solution which 
relates the bulk Lorentz factor $\Gamma(R)$ with the distance $R$ from the central source,
\begin{equation}
\Gamma(R)^2=\frac{17\, E_{\rm k}}{16\, \pi\, m_{\rm p}\, c^2\, n\, R^3}
\end{equation}
where $E_{\rm k}$ is the kinetic energy driving the expansion of the fireball into the interstellar medium of density $n$. The kinetic energy is related to 
the prompt emission $\gamma$--ray energy \eiso/$\eta \approx E_{\rm k}$ through the efficiency $\eta$. 

We can derive the time $t_{\rm v}$ when $\Gamma=1/\sin \theta_{\rm view}$ expressing through Eq.1 the distance $R_{\rm v}$ when this transition occurs:
\begin{equation}
R_{\rm v} = \left(\frac{17\, E_{\rm k}\, \sin^2 \theta_{\rm view}}{16\, \pi\, m_{\rm p}\, c^2\, n} \right)^{1/3}
\end{equation}

The corresponding time $t_{\rm v}$ can be derived by integrating: 
\begin{equation}
\int_{t_{\rm p}}^{t_{\rm v}} c\, dt = \int_{R_{\rm p}}^{R_{\rm v}}  \frac{1-\beta(r)\cos \theta_{\rm view}}{\beta(r)}\, dr
\end{equation}

The lower boundary of the integrals  ($R_{\rm p}$, $t_{\rm p}$) correspond to the distance/time from which the BM solution is valid. Although the 
transition from the coasting phase (when the fireball is moving with constant velocity $\Gamma_0$) to the deceleration phase is smooth (see e.g. 
Nava et al. 2013), we derive $R_{\rm p}$ by extrapolating backwards the BM solution to the end of the coasting phase, i.e. $R_{\rm p}=\left(17 E_{\rm k}/16\pi m_{\rm p} c^2 n \Gamma_0^2\right)^{1/3}$. For the typical parameters of our synthetic GRB population ($R_{\rm p}$, $t_{\rm p}$)$\ll$($R_{\rm v}$, $t_{\rm v}$) so 
that the estimate of $t_{\rm v}$ is dominated by $R_{\rm v}$. 

Following G13 (see also \S4), we assume  the ISM density $n$  uniformly distributed between 1 and 30 cm$^{-3}$ and a typical value of the radiative 
efficiency $\eta=20$\%. Fig.\ref{fg2} shows the differential (main panel) and cumulative (inset) distribution of the time of the peak of the OA, $t_{\rm v}$, 
of the simulated population of GRBs seen off--axis (solid blue line). The OA emission peaks on average a few years after the GRB event. 

However, the 
relevant timescale for OA studies and detection (\S5) is their duration because there is no starting reference time coincident with the GRB prompt emission 
(which is undetected for OA). The time when the OA emission starts to be visible $t_{\rm s}$, at a very low flux level, is when $\Gamma=1/\sin(\theta_{\rm view}-\theta_{\rm jet})$. Then the flux rises reaching a peak at $t_{\rm v}$ and decays afterwards in the same way as it would if seen from an on--axis observer. In order to 
define a duration, we consider the difference between the time when the jet becomes non--relativisic $t_{\rm NR}$ (i.e. when $\Gamma=1$, e.g. Livio 
and Waxman 2000) 
and the time when the OA starts to be visible ($t_{\rm s}$). 
The OA duration distribution is shown by the dashed (red) histograms in Fig.\ref{fg2} (differential and cumulative in the main 
plot and inset, respectively). The duration of the OA emission is slightly smaller than the typical timescales corresponding to the peak of the same 
emission confirming that OA are slow transients. 

\begin{center}
\begin{figure*}
\includegraphics[width=16.5cm,trim=40 10 20 10,clip=true]{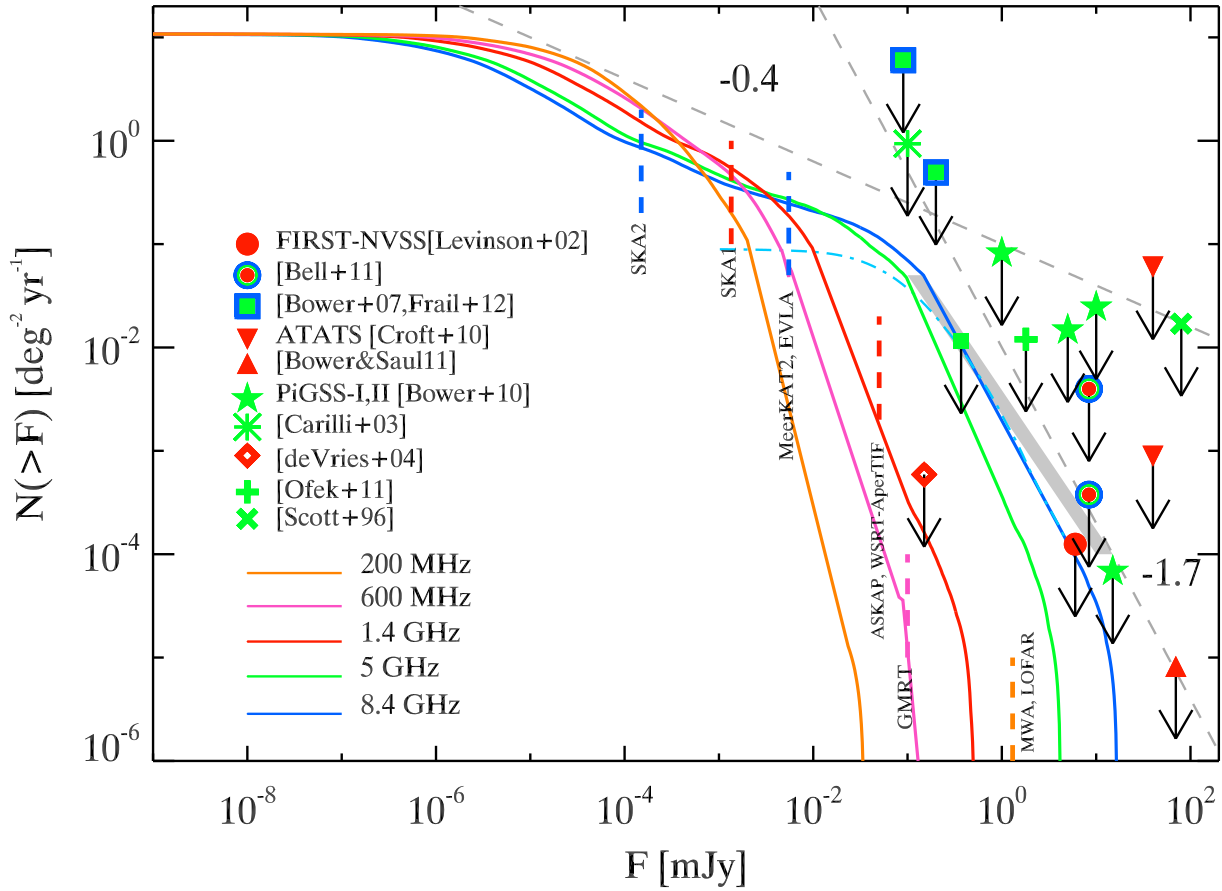}
\caption{Cumulative flux distribution of orphan afterglows at GHz and MHz observing frequencies (color codes as shown in the legend). The dashed 
lines (with slope $-1.7$ and $-0.4$) are shown for reference. The dot--dashed (cyan) line shows the flux distribution (at 8.4 GHz) of the subsample 
of GRBs with \thv$<$10$^{\circ}$ which determines the break. The current (3$\sigma$) upper limits on the rate of orphan afterglows detected in 
archival searches and radio surveys are shown (with colour codes corresponding to the observing frequencies) and the corresponding survey 
names/references are shown in the legend. The vertical dashed lines (colour codes corresponding to the sampling frequencies) represent the 
5$\sigma$ flux limits (Tab.1) that will be reached by current and future radio surveys (labelled with the corresponding instrument/survey name). The grey shaded thick line represents the predictions of Frail et al. (2012) at 8.4GHz.  }
\label{fg3}
\end{figure*}
\end{center}

\section{Orphan afterglow fluxes} 

To characterise the flux distribution of OA we have coupled the population synthesis code (G13) with an emission model for 
the afterglow. We use the afterglow Hydrodynamic Emission (HE) code of van Eerten \& MacFadyen (2012a, 2012b) obtaining a code (PSYCHE) 
which has already been used to study the radio emission properties of on--axis GRBs (Ghirlanda et al. 2013b). The HE code is based on a set of 
jet hydrodynamic 2D simulations that describes the evolution of the jet expansion into a constant density medium from the ultra--relativistic phase 
to the sub--relativistic one. This code assumes synchrotron emission with self absorption from a population of electrons accelerated at the shock 
front with a power law energy distribution with slope $p$. The fraction of the shock energy shared between electrons and magnetic field is parametrized 
by the $\epsilon_e$ and $\epsilon_B$ parameters.

In addition to the GRB parameters (redshift $z$, jet opening angle \th, viewing angle \thv, isotropic equivalent kinetic energy $E_{\rm k}$) PSYCHE 
requires to set $n$, $\epsilon_e$ and $\epsilon_B$. The values of these parameters will differ from burst to burst. Ghirlanda et al. (2013b) shows that 
with typical values of $\epsilon_e=2 \times 10^{-2}$ and $\epsilon_B=8\times10^{-3}$ and $p=2.5$ (see also Ghisellini et al. 2008), PSYCHE can 
reproduce the radio flux distribution of the BAT6 \sw\ sample. 

Through PSYCHE we can compute the flux density of the population of OA at typical characteristic frequencies: considering the current radio facilities, 
we choose three GHz frequencies (1.4, 5 and 8.4 GHz) and two MHz frequencies (200 and 600 MHz). Fig.\ref{fg3} shows the cumulative peak flux 
distribution of OA radio afterglows at these frequencies. 

The distributions of Fig.\ref{fg3} show the rate of OAs  in deg$^{-2}$ yr$^{-1}$. The bright end of the flux distribution extends to the mJy level at higher frequencies, although the rate of these bright events is very small. For instance, at 8.4 GHz (blue line in Fig.\ref{fg3}), there are  $\sim$ 2$\times 10^{-3}$ events deg$^{-2}$ yr$^{-1}$ brighter than 1 mJy.  
This rate is smaller by a factor 10--50 with respect to the predictions of Levinson et al. (2002) rescaled for our different assumptions. This could be due to the assumptions of that work (e.g. a unique beaming factor and energetic of GRBs and the description of the OA flux at the trans--relativistic transition). Our model instead assumes the proper distributions of the jet opening angles and GRB energetics (as derived through the population synthesis code of G13) and uses the HE code to describe the OA emission throughout the relativistic to non--relativistic phase.

The flux distributions shown in Fig.\ref{fg3} are consistent with a slope $-1.7$ at high fluxes and assume a flatter slope $-0.4$ at lower flux levels (grey dashed lines in Fig.\ref{fg3}). 
The bright end of the flux distribution (i.e. above 0.1 mJy) is consistent with the prediction of Frail et al. (2012) for the same flux interval (shown by the shaded grey thick line in Fig.~\ref{fg3}). 
However, we note that our code extends the flux energy range far below this limit where there is a considerable flattening of the flux distribution. This is a relevant point for the prediction of the rate of OA detectable by future deep radio surveys. 
The high end of the flux distribution is dominated by GRBs with small opening angles \th\ observed at small viewing angles \thv\ (low left corner of the \thv-\th\ plane in Fig.\ref{fg1}). In G13 we have shown that small \th\ correspond to GRBs with large \G, which have large energetics. The break of the flux distribution corresponds to a viewing angle \thv$\approx 10^{\circ}$ (as shown by the dot--dashed cyan line in Fig.\ref{fg3}). The slope of the flux distribution below the break is due to the superposition of the flux distributions of progressively more GRBs with larger \thv.

\section{Orphan Afterglows radio detection rates}

\subsection{Present limits}
Searches of  transients in radio archival observations or radio surveys (Levinson et al. 2002; Gal-Yam et al. 2006; Bannister et al. 2011; Bell et al. 2011; Bower and Saul 2011; Bower et al. 2007, 2010; Croft et al. 2010; Frail et al. 2012; Carilli et al. 2003; Matsumura et al. 2009; Lazio et al. 2010) set upper limits on the sky density of transients. 

\begin{center}
\begin{figure*}
\includegraphics[width=16.5cm,trim=40 10 20 10,clip=true]{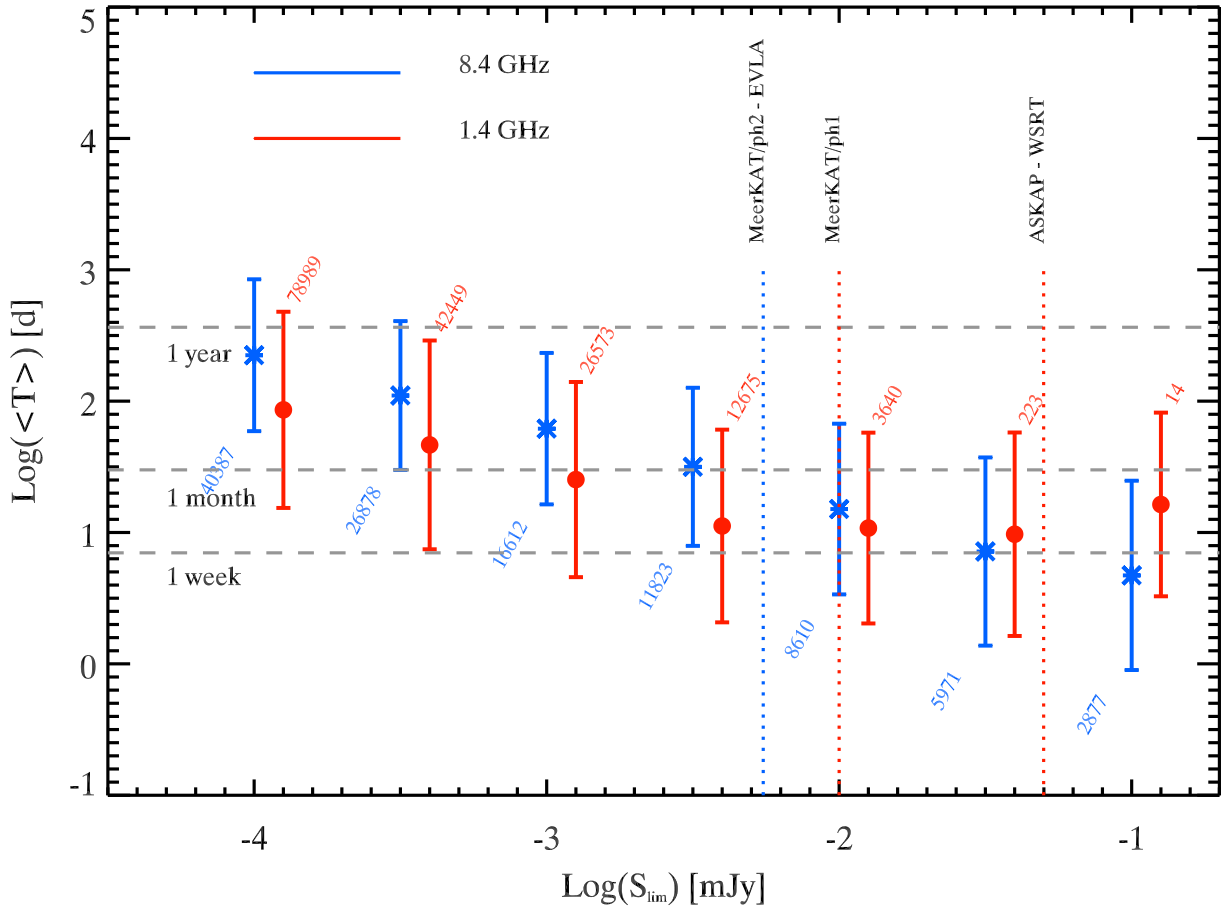}
\caption{Average OA duration above $S_{\rm lim}$ as a function of the survey limiting flux. Blue asterisks are for the 8.4 GHz and red circles for 1.4 GHz (the latter are slightly displaced along the abscissa for clarity). The reported numbers correspond to the total number of OA that are above $S_{\rm lim}$ at their peak (in units of yr$^{-1}$ all sky). The 5$\sigma$ limiting fluxes of the current and future surveys are reported. }
\label{fg4}
\end{figure*}
\end{center}

Detection of radio transients is, however, only the first step towards the identification of OA, because many other astronomical objects can produce radio transients (see e.g. Murphy et al. 2013 for a recent review) and the absence of any $\gamma$--ray trigger for OA prevents their classification as events related to GRBs seen off--axis. Of the nine candidate radio transients identified by comparing the NVSS (Condon et al. 1998) and FIRST (White et al. 1997) radio surveys at 1.4 GHz (Levinson et al. 2002), five were rejected as false triggers and two were classified  
as a radio SN and as an artifact in the data
by  Gal-Yam et al. (2006) through extensive follow up and multi wavelength observations. Similarly, the detection rates (10 transients at 8.4 and 4.8 GHz) originally reported by Bower et al. (2007) through the analysis of archival VLA observations, were later rejected by an independent analysis of the same data sets (Frail et al. 2012). The  fields of the VLA (at 1.4, 4.8 and 8.4 GHz), recently analysed by Bell et al. (2011), are distributed over more than 20 years and have typical separations of few days or a month but do not show any radio transient detected down to a limiting flux of 8 mJy. There were also works that used extensive observations of a single field (e.g. the archival VLA observations of the 3C 286 field - Bower \& Saul 2011 or the Lockman Hole - Carilli et al. 2003) all giving upper limits on the detection of radio transients at some flux level. Murphy et al. (2013)  and Bell et al. (2011) summarised these results.

Fig.\ref{fg3} shows the upper limits derived from Tab.3 of Murphy et al. (2013) at the corresponding flux limit of the survey. The limits on the radio transients density have been converted into detection rate limits considering the typical timescales of the surveys.  All the current upper limits are consistent with the flux distribution of the population of OA derived with PSYCHE. The different colours of the upper limits in Fig.\ref{fg3} correspond to the characteristic radio frequency of the survey and should be compared with the corresponding line (same colour coding) of the flux distribution. All the current limits correspond to relatively high flux levels, larger than 0.1 mJy. 

\subsection{Future surveys}

Considering the main future surveys that will be performed by the SKA and its pathfinders we report in Tab.\ref{tab1} the detection rates (col.4) expected considering a 5$\sigma$ flux limit (col.3). We are aware that the continuum sensitivity limits are still somewhat uncertain, but we adopt fiducial values from the available literature~\footnote{We adopted values from Fig. 1 of the SKA memo number SKA-TEL-SKO-DD-001, from Murphy et al. 2012, and from Macquart et al. 2010. We scaled the sensitivities to the same exposure duration of 12 hours, unless confusion is reached earlier.}. 
 
We also note that the future survey design is still on its way and the two leading parameters, i.e. the field of view and the sensitivity, should be considered in estimating the rate of AO detectable by a given survey at a given frequency. Here (Tab.\ref{tab1}) we give the detection rates in units of OA yr$^{-1}$ deg$^{-2}$ at fiducial $5\sigma$ sensitivity limits, so that if the sensitivity will remain almost unchanged the rate can be obtained multiplying for the field of view. If these numbers will change substantially for a given survey, Fig.\ref{fg3} should be used to derive the OA rate at a different sensitivity limit for surveys operating at different frequencies. 
As far as the MHz telescopes are concerned, we predict that the peak OA fluxes will be 50--100 times fainter than the $\sim$mJy sensitivity provided. 

\begin{table}
\caption{Detection rates of OA by future radio telescopes. For each survey the observing frequency (col.2) and the 5$\sigma$ sensitivity limit (col.3) is reported. The rates (col.4) are 
derived from the flux density distributions shown in Fig.\ref{fg3}.} 
\begin{center}\small
\begin{tabular}{cccc}
\hline \hline
Telescope name & $\nu$   & $S_{\rm lim}$ & Rate \\
	    		  &    [GHz]   &  [mJy]        	     & [deg$^{-2}$ yr$^{-1}$] \\
\hline
ASKAP		      		&   1.4			& 0.05		    &	$3\times10^{-3}$	\\
MeerKAT/Ph1		&	1.4	      		& 0.009		    &	$10^{-1}$	\\
MeerKAT/Ph2		&	8.4	      		& 0.006		    &	$3\times10^{-1}$	\\
SKA/Ph1			&	1.4			& 0.001		    &	$6\times10^{-1}$	\\
SKA/Ph2			& 	1.4(8.4)		& 0.00015	    &	$1.5$($2\times10^{-1}$)			\\
WSRT/AperTIF	&    1.4			& 0.05		    &	$3\times10^{-3}$	\\
EVLA				&	8.4			& 0.005		    &	$3\times10^{-1}$	\\
LOFAR				&	0.2			& 1.3		    &		...				\\
MWA				&	0.2	      	      & 1.1		    &		...				\\
GMRT				&	0.6		      &  0.1		    &	$10^{-5}$			\\
GMRT				&	1.4		      &  0.15		    &	$2\times10^{-4}$	\\	
\hline \hline
\end{tabular}\label{tab1}
\end{center}
\end{table}

In order to explore the population of OA at radio frequencies it is fundamental to go deeper than current limits. Considering the slopes of the flux distributions (Fig.\ref{fg3}) one should go deeper in sensitivity above the break of the flux distribution and instead consider a wider field of view at fluxes below the break in order to maximise the rate of detected OA.

Current searches for radio transients have used available radio observations (mostly archival) which are not homogeneously spaced in time. In general these searches are sensitive to radio transients which are present in some observation and then disappear (or the other way around). The timescale of the transients that are detectable is therefore related to the timescale separating subsequent observations (i.e. the survey cadence). Radio OA are  long lived transients and it is important to estimate how long they last. While we have defined their duration in \S3 considering the separation between two characteristic ``dynamical" timescales, here we want to describe the time interval during which the OA is above a certain flux threshold $S_{\rm lim}$ corresponding to a given survey limit. 
\begin{center}
\begin{figure}
\includegraphics[width=8.5cm,trim=40 10 20 10,clip=true]{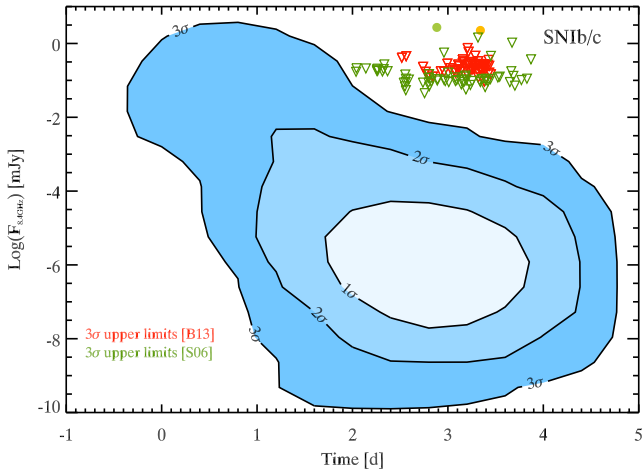}
\caption{
 Density contours (1,2 and 3$\sigma$ as labelled) representing the distribution of the flux (at 8.4GHz) of the OA population versus the time when their light curve peaks. The 3$\sigma$ upper limits of the SNIb/c observed in the radio band by Soderberg et al. 2006 (green triangles) and by Bietenholz et al. 2013 (red triangles) are shown.} 
Filled circles are the two detections at radio frequencies, i.e. SN 2001em and SN 2003gk.
\label{fg5}
\end{figure}
\end{center}

The number of OA above threshold and their average duration depends on $S_{\rm lim}$. Decreasing $S_{\rm lim}$ (i.e. for a deeper survey): (i) the fraction of OA that can be detected with flux $F\ge S_{\rm lim}$ increases and (ii) their ``duration above threshold"=T increases. We have computed for all the simulated OA the time interval during which their emission (at a given frequency) is above $S_{\rm lim}$. Fig.\ref{fg4} shows the average duration T above a certain $S_{\rm lim}$ as a function of $S_{\rm lim}$. The reported numbers are the all sky number of OA in units of yr$^{-1}$ which can be detected given that $S_{\rm lim}$. The average duration T above $S_{\rm lim}$ increases when deeper survey limits are considered. We note that future surveys (shown by the vertical dotted lines in Fig.\ref{fg4}) can detect a large number of OA per year (full sky) provided that their cadence is smaller or similar to the typical duration of OA (as shown in Fig.\ref{fg4}). For example, considering the ASKAP (or WSRT) flux limit of 50$\mu$Jy at 1.4 GHz, there are $\sim$125 OA yr$^{-1}$ on all the sky. Out of these, $\sim$0.1 yr$^{-1}$ could be detected in the VAST survey field of view of 30 deg$^2$. These sources could be detected as on--off transients in subsequent exposures separated at least by one week up to one month.

In G13 we derived that $\sim$0.3\% of SN Ib/c  can produce a GRB event. This percentage refers to all GRBs, i.e. those pointing to the Earth (i.e. detected as $\gamma$--ray events) and those pointing in other directions (detectable as OAs).  If SN Ib/c are GRBs oriented away from the observer line of sight, they should be detected at late times when the afterglow emission has decelerated enough to encompass, with its beaming angle, the observer viewing angle \thv.
Soderberg et al. (2006 - S06) and Bietenholz et al. (2013 - B13) performed radio surveys of a sample of SN Ib/c, the putative progenitors of long--duration GRBs. 
Since their combined sample consists of 112 SN, according to the finding of G13 we should expect that $\sim$0.34 SN Ib/c of their sample can harbour a GRB jet, i.e. we would expect no detection. They observed these SN\ae~at late times (years after the explosion) at 8.4 GHz. 
S06 and B13 report indeed upper limits on the late time radio flux of the monitored sources  with the exception of SN2001em (see S06), and SN2003gk (see B13), which are in fact detected. 
Nonetheless, further monitoring of these two events in the radio band and through VLBI observations, excluded that they produced a relativistic jet. We show in Fig.\ref{fg5} the upper limits on the 8.4 GHz flux of the SN Ib/c observed by S06 and B13 which are all consistent with the density contour of the distribution of the simulated population of OA. The 1, 2, and 3$\sigma$ contours represent the boundary containing respectively 68.2\%, 95.4\%, and 99.7\% of the points distribution in the plane $F_{\rm peak}$ -- \tpeak.


\section{Summary and Discussion}

Orphan afterglows are GRBs whose emission is detectable only during the afterglow phase (at optical/NIR and radio frequencies). Their prompt $\gamma$--ray emission is unobservable because the viewing angle \thv\ is larger than the jet opening angle \th\ (off--axis GRBs). In these events the afterglow emission becomes observable when the bulk Lorentz factor, which is decreasing during the afterglow phase, becomes $\Gamma \sim 1/$\thv. After this time, 
which represents the peak of the OA light curve, the emission is similar to that for an observer within the jet opening angle. 

OA make up a majority of the population of GRBs. However, none have been observed so far, do to their lack of a prompt emission trigger. Their detection is possible as transients in deep/wide field surveys. However, so far no detection of OA has been confirmed by searches in archival optical/radio observations. In current and future surveys OA might  represent a considerable fraction of detected transients. 

In this paper we have used the results of a population synthesis code for GRBs (G13) that simulates the entire population of GRBs including off-axis events and is anchored to reproduce some observational constraints of the population of  GRBs detected \fe\ and \cgro\ with particular emphasis on the constraints given by the BAT6 complete \sw\ sample (Salvaterra et al. 2012). 

We have explored the properties of the population of off--axis GRBs (see Fig.\ref{fg1} - black symbols) in terms of their radio emission. We have computed the radio flux density of the OA population (representing $\sim$97\% of the entire GRB simulated population) at the time when the OA light curve reaches its peak (Fig.\ref{fg2}) which is of the order of few years after the prompt trigger. However, the lack of any prompt emission (i.e. $\gamma$--ray trigger) in OA, requires to compute the timescale of their duration which, given the typical rise/decay long--term evolution of the afterglow flux, can be of the same order of the peak time. This suggest that OA in the radio band should be slow transients. 

We have constructed the cumulative flux distribution at different radio frequencies (GHz and MHz) that shows a high flux tail with a slope consistent with $-1.7$ (Fig.\ref{fg3}) and a break at a frequency dependent flux below which the slope becomes flatter ($-0.4$). This is due to the combination of the \th\ distribution of the population of simulated bursts with the viewing angle \thv\ probability function: at approximately 10$^{\circ}$ the product of the \th\ log--normal distribution (resulting from the population code of G13) and of the probability density of \thv\ is maximised. This accounts for the slope change of the flux distributions in Fig.\ref{fg3}, the break of the flux distribution is shifted to lower fluxes at lower frequencies because the radio emission is in the self absorbed regime of the synchrotron spectrum. 

In general, from Fig.\ref{fg3} we note that very bright OA with flux at the 1 mJy level are rare. This is the central flux of the current radio searches/surveys that have been searching for OAs. Our population is consistent with the limits given by these surveys. Totani \& Panaitescu (2002) derived the flux distribution of orphan afterglows based on 10 bright GRB afterglows. Our estimates are consistent in the bright flux end with their but we predict a lower number of OA at low fluxes (below the break of our flux distribution). This is due to the modelling, in our case, of the GRB jet opening angle distribution (G13).  

In G13 we derived that $\sim$0.3\% of SN Ib/c may harbour a GRB, i.e. in other words the great majority of the putative progenitors of GRBs do not produce a relativistic jet. When compared to the combined samples of S06 and B13, which comprise 112 SN Ib/c, there were indeed only two detections and we statistically expect none of them to be an off-axis GRB. Both SN 2001em and SN 2003gk were successively showed not to expand at relativistic velocity, arguing against their being genuine OAs. Finally, we showed (in Fig.~\ref{fg5}) that all upper limits on SN Ib/c radio emission at late times are indeed consistent with the distribution of GRB OAs in the plane $F_{\rm peak}$ -- \tpeak. 
 
\acknowledgements{We thank the PRIN/INAF C41J12000020005 for financial support. DB and TM acknowledge the support of the Australian Research Council through grant DP110102034. The Centre for All-sky Astrophysics is an Australian Research Council Centre of Excellence, funded by grant CE110001020. The anonymous referee is acknowledged for his/her useful comments. }

\end{document}